\documentclass[aip,jap,reprint,amsmath,amssymb]{revtex4-1}
\usepackage{graphicx}
\usepackage{dcolumn}
\usepackage{bm}
\usepackage{changes}
\usepackage{here}
\usepackage{physics}

\usepackage{filecontents}

\begin{document}


\title{Laser-to-droplet alignment sensitivity relevant for laser-produced plasma sources of extreme ultraviolet light} 

\author{Sten~A.~Reijers}
\affiliation{Physics of Fluids Group, Faculty of Science and Technology, MESA+Institute, University of Twente, P.O. Box 217, 7500 AE Enschede, The Netherlands}

\author{Dmitry~Kurilovich}
\affiliation{Advanced Research Center for Nanolithography (ARCNL), Science Park 110, 1098 XG Amsterdam, The Netherlands}
\affiliation{Department of Physics and Astronomy, and LaserLaB, Vrije Universiteit Amsterdam, De Boelelaan 1081, 1081 HV Amsterdam, The Netherlands}

\author{Francesco~Torretti}
\affiliation{Advanced Research Center for Nanolithography (ARCNL), Science Park 110, 1098 XG Amsterdam, The Netherlands}
\affiliation{Department of Physics and Astronomy, and LaserLaB, Vrije Universiteit Amsterdam, De Boelelaan 1081, 1081 HV Amsterdam, The Netherlands}

\author{Hanneke~Gelderblom}
\affiliation{Physics of Fluids Group, Faculty of Science and Technology, MESA+Institute, University of Twente, P.O. Box 217, 7500 AE Enschede, The Netherlands}

\author{Oscar~O.~Versolato}
\email{o.versolato@arcnl.nl}
\affiliation{Advanced Research Center for Nanolithography (ARCNL), Science Park 110, 1098 XG Amsterdam, The Netherlands}

\date{\today}

\begin{abstract}
We present and experimentally validate a model describing the sensitivity of the tilt angle, expansion and propulsion velocity of a tin micro-droplet irradiated by a 1\,$\mu m$ Nd:YAG laser pulse to its relative alignment. This sensitivity is particularly relevant in industrial plasma sources of extreme ultraviolet light for nanolithographic applications. Our model has but a single parameter: the dimensionless ratio of the laser spot size to the effective size of the droplet, which is related to the position of the plasma critical density surface. Our model enables the development of straightforward scaling arguments in turn enabling precise control the alignment sensitivity.
\end{abstract}

\maketitle

\section{\label{sec:intro}Introduction}
\begin{figure}[b]
\includegraphics[scale=1]{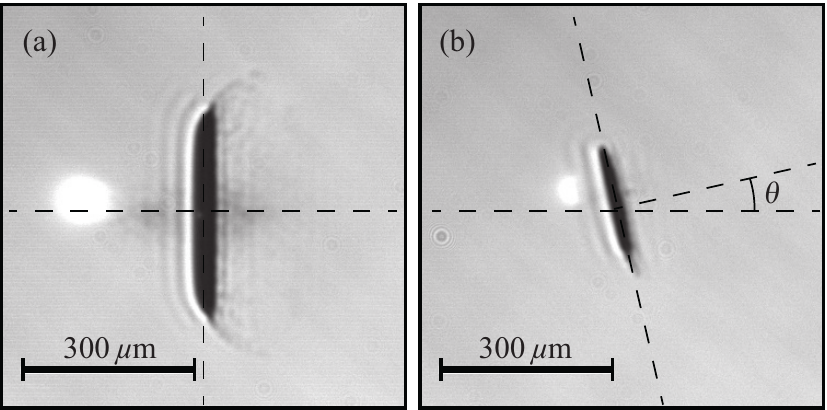}
\caption{\label{fig:shadow} Side-view shadowgraphy images taken 2\,$\mu$s after a 40\,mJ pulse with 115\,$\mu$m diameter (FWHM) spot impinged on a droplet from the left at (a) optimal alignment and at (b) a misalignment of $\Delta x \approx 100$\,$\mu$m leading to a tilt angle $\theta$, accompanied by a decrease in total propulsion velocity and radial expansion. The white glow visible at the original droplet position is plasma light, where the deformed droplet has moved further downwards and sideways (see main text).}
\end{figure}

Microdroplets of liquid tin are used to create extreme ultraviolet light (EUV) for next-generation nanolithography\cite{Benschop2008,Banine2011, Fomenkov2017, Kawasuji2017} that is currently being introduced in high-volume manufacturing. These droplets, several 10\,$\mu$m in diameter, serve as mass-limited targets \cite{Fujioka2008,OSullivan2015} for creating a laser-produced plasma (LPP) in EUV light sources. In such machines, a prepulse laser beam hits a tin droplet to obtain an extended disk-like target\cite{Gelderblom2016,Kurilovich2016} that increases coupling with the next pulse. Subsequently, the target is irradiated by a focused nanosecond-pulse laser at intensities that lead the creation of a high-density plasma\cite{Banine2011,Tomita2015,OSullivan2015}. Line emission from electron-impact-excited highly charged tin ions in the plasma provides the EUV light peaking at 13.5\,nm \cite{Banine2011,OSullivan2015}. Maximizing the conversion efficiency (CE) of laser light into the required EUV light of such sources requires a careful control over the target shape. This shape is very sensitive to the precise alignment of the prepulse laser to the initially spherical droplet \cite{nakamura2008ablation,hudgins2016neutral}. Any deviation from the optimum location for laser impact will produce a suboptimal target tilt and decrease the target radial expansion, as shown in Fig.~\ref{fig:shadow}. Moreover, reflections of the laser light from the incorrectly tilted surface of the target back towards the laser itself may be detrimental to laser operation stability. In spite of its obvious relevance, the precise relation between such laser-to-droplet (L2D) alignment and the resulting target tilt has so far remained poorly explored aside from activities by Tsygvintsev et al. \cite{Tsygvintsev2014} and the recent work by Hudgins et al. \cite{hudgins2016neutral} who combine modeling with experimental efforts. Due to experimental constraints, however, their model predictions for the tilt sensitivity could not be validated under conditions of controlled misalignment.

Here, we present and experimentally validate an intuitive model describing L2D alignment sensitivity. The model is based on a single parameter: the dimensionless ratio of the laser spot size to the effective size of the droplet target, which we relate to the position of the plasma critical density surface \cite{Basko2015,5910405,Sizyuk2013} and sets the typical length scale for the problem. Our model enables the development of straightforward scaling arguments, which in turn enable the minimization of detrimental alignment sensitivity.
We focus our studies on industrially relevant 1-$\mu$m-wavelength Nd:YAG laser pulses and experimentally validate our model predictions for two different laser spot sizes over a wide range of laser pulse energies. Furthermore, we apply the validated model to predict sensitivities for several practical cases that are immediately relevant for current state-of-the-art industrial droplet-based EUV light sources.

\section{\label{sec:expt}Experiment}

Our experimental setup has previously been described in detail\cite{Kurilovich2016}. For clarity, the most important characteristics are repeated here. A droplet generator was operated in a high-vacuum chamber ($\sim$\,$10^{-7}$ mbar) and held at constant temperature of $\sim$\,$260^{\circ}$\,C. The nozzle produced a 13.4\,kHz train of $\sim$\,$43$\,$\mu$m diameter (radius $R_0 \approx 21.5$\,$\mu$m) droplets of 99.995\% purity tin. The droplets were irradiated by 1064-nm wavelength, 10\,ns (full-width-at-half-maximum: FWHM) long pulses of a Nd:YAG laser operated at 10\,Hz repetition rate.

The laser pulse energies were varied between $\sim$\,1 and $\sim$\,395\,mJ. The laser beam was focused down to a 115\,$\mu$m or 60\,$\mu$m diameter (FWHM) Gaussian spot on the droplet and was circularly polarized. While significant astigmatism was apparent for a tighter focus, the part of the beam intersecting with the droplet could still be well described by a 60\,$\mu$m-diameter Gaussian function. Conversely, the 115\,$\mu$m focus produced a circularly symmetric beam spot. In order to capture the dynamics of the expanding droplets two shadowgraphy systems based on 850-nm-wavelength, 15-ns pulsed laser diodes and long-distance microscopes coupled with charge-coupled-device cameras were used. The two systems provide a ``front'' view and a side view, $30^{\circ}$ and $90^{\circ}$ with respect to the laser propagation direction. By varying the time delay between the plasma-generating laser pulse and shadowgraphy pulses, a sequence of images was recorded. The images obtained from the side were processed using an image analysis program that tracks the center-of-mass displacement and the size of the expanding droplet as well as the target tilt angle. In order to introduce controlled misalignment, the timing of the Nd:YAG laser pulse was varied around the established time for optimal alignment of laser on droplet, see Fig.~\ref{fig:shadow}. Since the droplets vertical velocity $U_x$ stayed constant ($\sim$\,12\,m/s), this ``mistiming'' $\Delta t$ resulted in a laser impact off-centered by a distance $\Delta x$, translating the initial spherical droplet into a tilted disk. The final droplet shape is the result of complex force interplay and its study is left for future work. The velocity $U_x$ was obtained by processing the front-view images containing two or more droplets. The target tilt angle, defined as the angle between the target normal and the laser beam propagation direction (see Fig.~\ref{fig:shadow}b), was mapped as a function of misalignment, with $\Delta x=U_x \Delta t$, for different laser pulse energies. At the same time, we tracked the propulsion velocity $u_z$ along the laser light propagation direction. 
We obtained an estimate of the droplet expansion velocity $\dot R$ by measuring its radius (i.e. the maximum droplet radius along the tilted axis) shortly (1\,$\mu$s) after the laser pulse impact and assuming a linear expansion on this short timescale, such that $\dot R \approx (R(1\,\text{$\mu$s})-R_0)/(1\,\text{$\mu$s})$.

\section{\label{sec:theo} Model}
\begin{figure}[t!]
\includegraphics[scale=1]{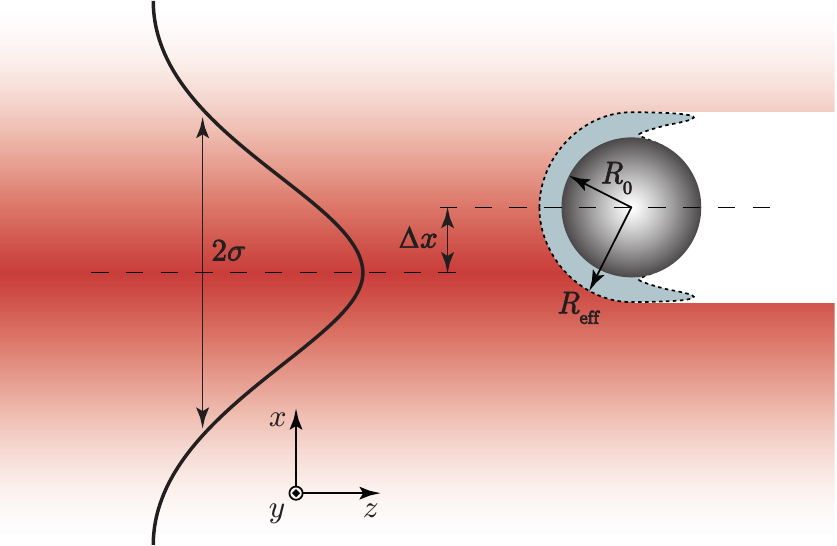}
\caption{\label{fig:schematics} Schematic cross-section of the problem:  a droplet with radius $R_0$ is hit by a Gaussian-shaped intensity-profile laser pulse with size $\sigma$ from the left. The laser hits the droplet off-center with an axial offset $\Delta x$ creating plasma (light gray area) surrounding the droplet on the illuminated side (dashed line), thereby creating an effective radius $R_{\textrm{eff}}$.}
\end{figure}

In this section, we outline an intuitive model for the tilt angle $\theta_{\text{tilt}}$ (see Fig.\,\ref{fig:shadow}b) in terms of the axial misalignment and the laser beam width. Figure~\ref{fig:schematics} shows a schematic overview of the problem. A droplet with radius $R_0$ is hit by a Gaussian off-centered laser pulse having an axial offset $\Delta x$ and beam width $\sigma$, which relates to the full-width at half-maximum of the laser beam according to the usual convention $\sigma = \text{FWHM}/(2\sqrt{2\ln2})$. The target tilt angle can be deduced from the direction of the target's center-of-mass motion. We derive the target's center-of-mass motion from the plasma pressure distribution on the surface of the target where the laser energy is absorbed. This pressure distribution is directly related to the spatial intensity profile of the laser. Below we detail these steps.

During the laser impact, a small liquid layer is ablated and forms a plasma cloud on the illuminated side of the droplet, see Fig. \ref{fig:schematics}. The time scale of this plasma generation is typically much shorter than the laser pulse. As a result, most of the laser energy is absorbed just before the plasma critical density surface\cite{Basko2015} where the plasma electron frequency equals that of the laser light and no further light penetration is possible. This gives the system an effective radius $R_{\text{eff}}\geq R_0$, see Fig.~\ref{fig:schematics}. The laser intensity profile $I$ at $R_{\text{eff}}$ is given by a projection of part of the Gaussian laser beam profile onto the laser-facing hemisphere with radius $R_{\text{eff}}$ and an offset $\Delta x$. It is insightful to formulate this intensity profile in terms of two dimensionless parameters: the dimensionless laser beam width $\alpha =\sigma/ R_{\text{eff}}$ and the dimensionless laser offset  $\beta = \Delta x / R_{\text{eff}}$, which gives in spherical coordinates
\begin{align}
I\textrm(\theta,\phi) &\sim \exp\left(-\left[\frac{\sin(\theta)^2 - 2 \beta \sin(\theta)\cos(\phi) + \beta^2}{2\alpha^2}\right]\right) \nonumber\\
&\times H(\pi/2 - \theta)\cos(\theta),
\label{eq:laserintensityatcriticalsurface}
\end{align}
where $\theta$ and $\phi$ are the polar angle and azimuthal angle respectively, and $H$ is the Heaviside step function to limit the intensity profile to the illuminated side of the droplet. Note that an extra $\cos(\theta)$ term is included in Eq.~\eqref{eq:laserintensityatcriticalsurface} to account for the projection of the beam onto the effective radius, which is assumed to be locally spherical and of constant size.


With the intensity profile at $R_{\text{eff}}$ at hand, we now obtain an expression for the plasma ablation pressure $p_{\text{a}}$ acting on this surface. A power-law dependence between this pressure and the impinging laser pulse energy has been established in previous work \cite{Kurilovich2016,hudgins2016neutral,Kurilovich2018}, and experimentally shown to be valid to excellent accuracy over three orders of magnitude in laser energy \cite{Kurilovich2016,Kurilovich2018}, 
\begin{align}
p_{\text{a}} \propto \left(E-E_0 \right)^{\delta},
\label{eq:criticalsurfacescaling}
\end{align}
where $E \propto I$ is the fraction of the laser pulse energy that is intercepted by the liquid target. 
In the remainder of this work, we take $\delta=0.6$ in accordance to Ref.~[\onlinecite{Basko2015}]. Furthermore we neglect the small offset energy $E_0$, since typically $E_0 \ll E$ (see Ref.~[\onlinecite{Kurilovich2016}]) and use Eqs.~\eqref{eq:laserintensityatcriticalsurface} and (\ref{eq:criticalsurfacescaling}) to find the local pressure $p_\text{a}(\theta,\phi)$ acting on the target's effective surface.

The center-of-mass velocity of the target is then obtained from $p_{\text{a}}$. To this end, we consider the droplet and its surrounding plasma cloud as a single body with radius $R_{\text{eff}}$ that is subjected to a local pressure distribution $p_\text{a}(\theta,\phi)$. Note that by doing so we neglect the exchange of momentum between this plasma cloud and the droplet. Furthermore, we assume that the liquid body does not deform on the time scale of the pressure pulse, which is justified since the timescale of deformation is typically much longer \cite{Gelderblom2016, Kurilovich2016}. The center-of-mass velocity of the body is then given by

\begin{align}
\label{eq:centerofmassequations}
u_x(\alpha,\beta) & \sim \nonumber \\
&\int_0^{\pi}\int_0^{2\pi} p_{\text{a}}(\theta,\phi,\alpha,\beta) \sin(\theta)\sin(\theta)\cos(\phi)\,\mathrm{d}\theta\,\mathrm{d}\phi,\nonumber \\
u_y(\alpha,\beta) & \sim \nonumber \\
&\int_0^{\pi}\int_0^{2\pi} p_{\text{a}}(\theta,\phi,\alpha,\beta) \sin(\theta)\sin(\theta)\sin(\phi)\,\mathrm{d}\theta\,\mathrm{d}\phi,\nonumber \\ 
u_z(\alpha,\beta) & \sim \nonumber \\
&\int_0^{\pi}\int_0^{2\pi} p_{\text{a}}(\theta,\phi,\alpha,\beta) \sin(\theta)\cos(\theta)\,\mathrm{d}\theta\,\mathrm{d}\phi,
\end{align}
where the coordinates $x$, $y$, $z$ are defined in Fig.~\ref{fig:schematics}. 

From the direction of the center-of-mass motion, one can now deduce the target tilt angle. Note that since a pressure always acts perpendicular to the surface it can never induce any rotation of the body.  Therefore the target tilt is a result of the preferred expansion direction of the liquid, which by definition is perpendicular to the direction of the center-of-mass motion. As a result, the tilt angle in radians is given by
\begin{align}
\theta_{\text{tilt}}(\alpha,\beta) = \arctan(u_x(\alpha,\beta)/u_z(\alpha,\beta)),
\label{eq:tiltangledefinition}
\end{align}
which needs to be evaluated numerically. Our approach differs from the one presented in Ref.~[\onlinecite{hudgins2016neutral}] as we include the full pressure distribution on the surface of the droplet, see  Eqs.~\eqref{eq:centerofmassequations}. Comparing the two models, our approach consistently yields significantly lower tilt angle sensitivities.

The target tilt angle sensitivity around zero misalignment is of interest for certain industrial applications \cite{rafac2016}. To this end we define the tilt angle sensitivity around $\beta=0$ as
\begin{align}
f(\alpha) =  \frac{\partial \theta_{\text{tilt}}(\alpha,\beta)}{\partial \beta}\bigg|_{\beta=0}, 
\label{eq:tiltangletheoretical}
\end{align}
which enables a straightforward inspection of the influence of the dimensionless beam width $\alpha$. 
The tilt angle sensitivity as expressed by Eq.~\eqref{eq:tiltangletheoretical} can be approximated analytically by expanding Eq.~\eqref{eq:laserintensityatcriticalsurface} and \eqref{eq:tiltangledefinition} up to $\mathcal{O}(\beta^2)$ and results in
\begin{align}
f(\alpha) =  \Re\left(\frac{u_x(\alpha,1)}{u_z(\alpha,0)}\right).
\label{eq:tiltangletheoreticalfull}
\end{align}
The full complex expression is given in its explicit form in the Appendix. The actual tilt angle in this approximation is given by
\begin{align}
\theta_{\text{tilt}}(\alpha,\beta)  \approx \frac{\partial \theta_{\text{tilt}}(\alpha,\beta)}{\partial \beta}\bigg|_{\beta=0} \beta. 
\label{eq:tiltangletheoreticalexpaned}
\end{align}

Another important industrially relevant parameter is the radial expansion velocity $\dot R$ as a function of the misalignment. In Fig.~\ref{fig:shadow} we observe that laser misalignment not only induces a target tilt, but also significantly decreases the expansion velocity. We now employ our basic model to obtain a first-order estimate of this reduced expansion. When the laser beam is misaligned with respect to the droplet, the laser intensity absorbed by the droplet decreases. As a consequence, both the center-of-mass speed and the expansion rate of the target decrease. The partitioning of kinetic energy between propulsion (center-of-mass motion) and expansion is set by the laser beam (or pressure) profile acting on the droplet as detailed in Ref.~[\onlinecite{Gelderblom2016}].
To obtain an intuitive, first-order estimate of the target expansion velocity, we assume that this energy partitioning remains fixed and is not influenced by the misalignment or laser beam energy, such that
\begin{align}
\frac{\dot{R}(\Delta x/R_0)}{ \dot{R}(0)} \sim \frac{U_{\text{cm}}(\Delta x/R_0)}{U_{\text{cm}}(0)}.
\label{expansion}
\end{align}
Here, the left-hand-side is the expansion velocity as function of the misalignment normalized by the expansion velocity at zero misalignment, and the right-hand-side is the center-of-mass velocity as function of the misalignment normalized by the center-of-mass velocity at zero misalignment. 

\section{Results} \label{sec:RES}
\begin{figure}[H]
\includegraphics[width=8.5cm]{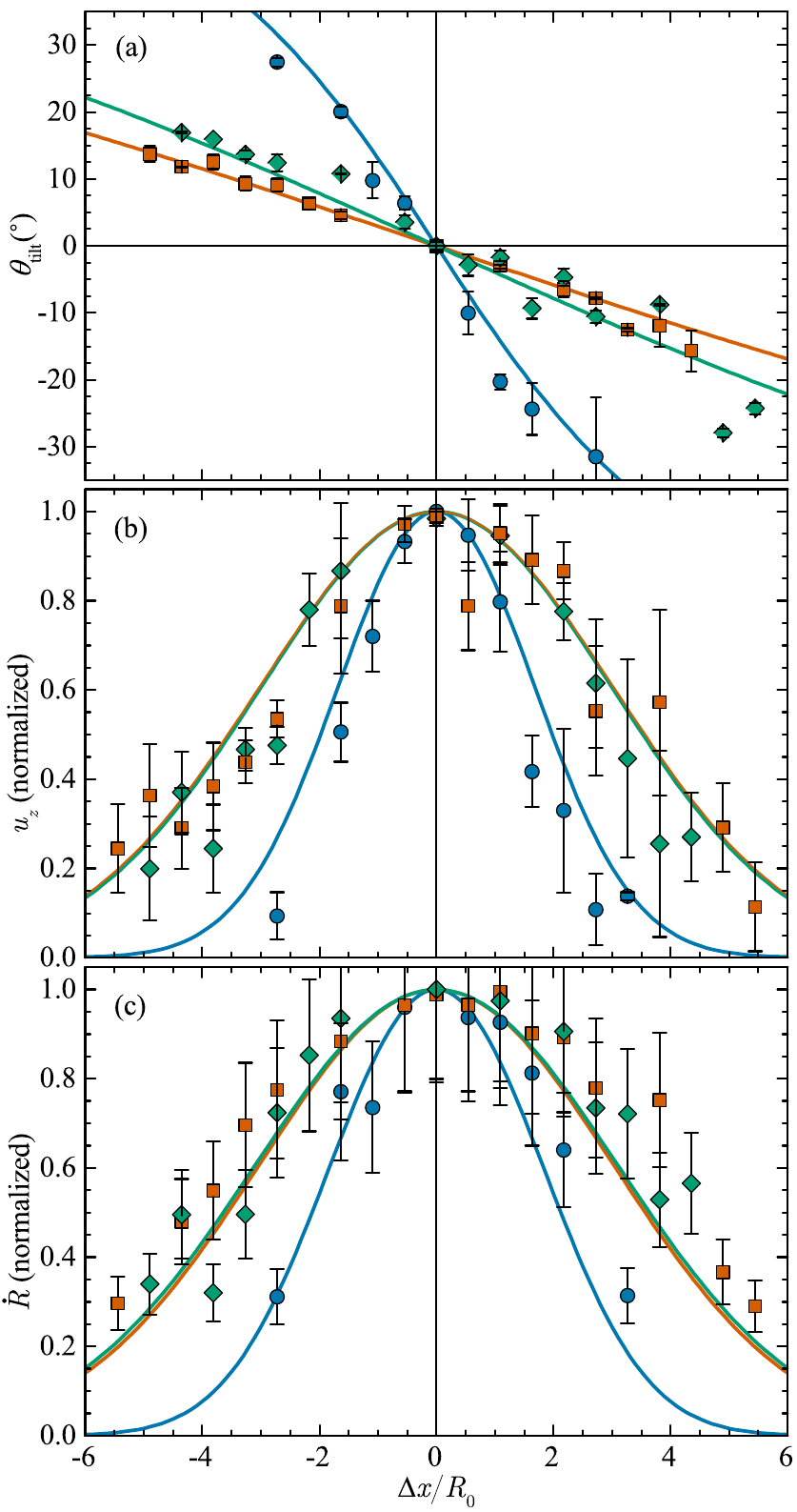}
\caption{\label{fig:results} (a) Tilt angle of the target at various misalignments. Data for the 115-$\mu$m focus with energies below (red squares) and above (green diamonds) 40\,mJ have been grouped. Blue circles represent averages of all data for the 60-$\mu$m focus case. Red, green, and blue solid lines depict numerical predictions of target tilt with $\alpha = 2.3$, $1.7$ and $0.86$ respectively. (b) Droplet propulsion velocity $u_z$ in the laser beam propagation direction. The red and green solid lines fully overlap. (c) Normalized droplet expansion velocity $\dot R$. In (b) and (c), the velocities have been normalized to the maximum value of each data set.}
\end{figure}

The experimental results for the tilt angle, the $z$-component of the center-of-mass velocity $u_z$ and the radial expansion velocity $\dot R$ are shown in Fig.~\ref{fig:results} as a function of misalignment $\Delta x/R_0$. The error bars represent the standard deviation of the measurements where available, otherwise a conservative value of twice the overall average error was used. In Fig.~\ref{fig:results}c the error bars are conservatively set at 20\%. We classify three different groups of experimental data. The first data set is obtained for a 115-$\mu$m focus with energies of $\sim$\,40 to $\sim$\,395\,mJ (green diamonds). The second group represents the experimental data for the same focus spot size, but with laser energies between $\sim$\,5 and $\sim$\,25\,mJ (red squares). The third group consists of all data having a 60-$\mu$m focus spot size with laser pulse energies between $\sim$\,1 and $\sim$\,95\,mJ (blue circles). In the top panel, we observe that the tilt angle monotonically increases with the misalignment. A small but significant influence of the laser pulse energy on the sensitivity is observed in the grouped data of the large focus spot size (green diamonds versus red squares) but this could not be further proven with any measure of significance for the individual (i.e. for single laser pulse energies) data sets. The small focus spot size (blue circles) results in a stronger tilt angle sensitivity to misalignment and show some signs of saturation at large misalignment values.  
In the center panel, we see that the normalized $z$-velocity $u_z$ distribution has a typical bell shape, showing no significant influence of the laser pulse energy in the grouped data of the large focus spot size. The smaller spot size results in a more sharply peaked distribution. 
In the bottom panel, we observe that the normalized radial expansion velocity $\dot R$ also has a typical bell shape, which appears to be of slightly larger width than the $u_z$ distribution. The data also shows no significant influence of the laser pulse energy in the grouped data of the large focus spot size.  Again, the smaller spot size results in a more sharply peaked distribution. 

To compare the experimental results in Fig. \ref{fig:results} to the model described above, we numerically evaluate Eq.~\eqref{eq:tiltangledefinition} using a local adaptive solver\cite{mathematica}. For each experimental case we determine $\alpha$ based on the focal spot sizes mentioned above and the effective radius $R_{\text{eff}}$, which we relate to the location of the critical plasma surface. We obtain the location of the critical surface for a Nd:YAG laser pulse on tin droplets from 2D radiation-hydrodynamic simulations \cite{Basko2015}. In that work, the distance from a tin droplet (at $R_0 = 15\,\mu$m) to the critical surface is evaluated to be $d_{\text{crit}}\approx 8$\,$\mu$m for a Nd:YAG laser pulse. By assuming the same position of the critical surface with respect to the droplet surface in our experimental case with slightly bigger droplets ($R_0 = 21$\,$\mu$m), we obtain $R_{\text{eff}} = R_0 + d_{\text{crit}} = 29.5$\,$\mu$m, and hence $\alpha_{115} = 1.7$ (115-$\mu$m focus, green diamonds) and $\alpha_{60}=0.86$ (60-$\mu$m, blue circles).
The corresponding sensitivity curve is in excellent agreement with the experimental data (see solid lines in fig.\,\ref{fig:results}).

For laser energies lower than 40\,mJ and a large focus spot size (red squares), the experimental data are found to be well described by the model if we set $\tilde{\alpha}_{115} = 2.3$, i.e., with the effective radius $R_{\text{eff}} \approx R_0$. This observation could be explained by the reasoning that for such for low-energy and broad focus pulses the plasma is not fully developed and the critical surface is situated very close to the droplet surface ($d_{\text{crit}} \approx 0$). Unfortuantely, no simulation data is available in this regime to support this claim. Furthermore, the difference between $\alpha_{115}$ and $\tilde{\alpha}_{115}$ could also be the result of different plasma pressure distributions \cite{Kurilovich2018} or additional dynamical, time-dependent effects that we do not consider here.  

Using the above mentioned values for $\alpha$, we observe that all theoretical curves show good agreement with the experimental data for both the tilt angle and the $z$-velocity $u_z$. Following the simple approximation given by Eq.~\eqref{expansion}, we compare the normalized, experimental radial expansion velocity to the theory predictions in Fig.~\ref{fig:results}. We find reasonable agreement between data and our model especially considering the simplifications involved and the experimental uncertainty in determining $\dot R$.
\begin{figure}[b]
\includegraphics[width=\linewidth]{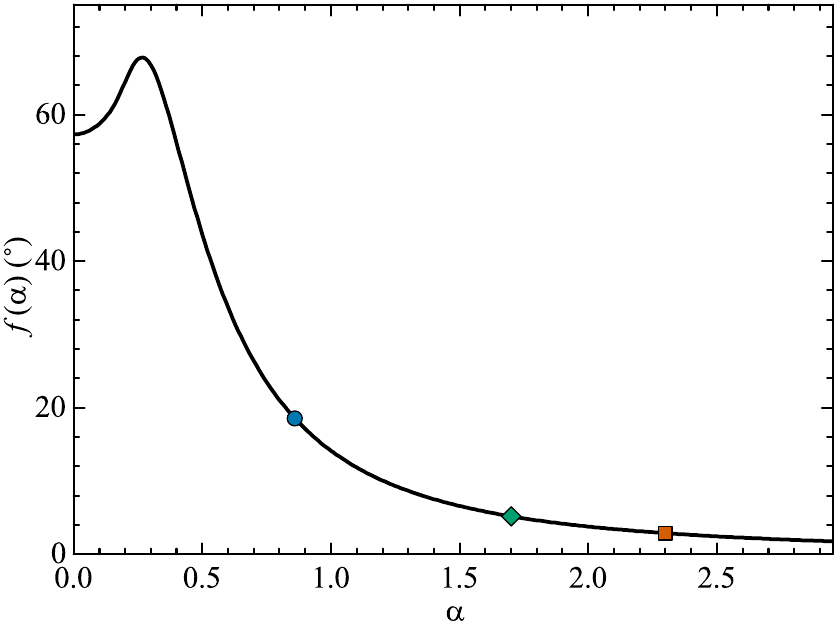}
\caption{\label{fig:tiltanglesensitivity} The theoretical tilt angle sensitivity $f(\alpha)$ (in degrees) as function of the dimensionless beam width $\alpha$ around zero misalignment ($\beta = 0$).  The black curve is the analytically calculated tilt angle sensitivity from Eq.~\eqref{eq:tiltangletheoretical}, with $\delta_p = 0.6$. The red square, green diamond and blue circle represent the values corresponding to the experimental conditions (see Section~\ref{sec:RES}, color coding as in Fig.~\ref{fig:results}).}
\end{figure}

Next, we analyze the model prediction for the tilt angle sensitivity around zero misalignment  ($\beta = 0$). Figure \ref{fig:tiltanglesensitivity} shows the numerical curve (solid line) resulting from Eq.~\eqref{eq:tiltangletheoretical} with the dimensionless beam widths corresponding to the experimental operating conditions (see Section~\ref{sec:RES}). The theoretical curve clearly shows that $f(\alpha)$ first increases, from unity with $\alpha$ peaking near $\alpha \approx 0.3$ after which a monotonic decrease is apparent. The sensitivity rises for larger droplets, keeping a constant beam width. Reversely, a smaller beam for a given effective radius will translate a large sensitivity to its alignment. In the limit $\alpha\to\infty$ one would illuminate the droplet with a flat-top beam of infinite width and would be completely insensitive to misalignment. In the limit $\alpha\to 0$ (i.e. a delta peak) the sensitivity decreases and eventually saturates to unity. In this limit the center-of-mass velocity $u_x$ and $u_z$ decay to zero equally fast as all energy is used to deform the droplet rather then to move its center-of-mass\cite{Gelderblom2016}. Hence the ratio $u_x/u_z$ becomes meaningless and one needs to reconsider the definition of the tilt angle. The maximum in $f(\alpha)$ is caused by a maximum in $u_x(\alpha,1)$, see Eq.~\eqref{eq:tiltangletheoreticalfull}. As $\alpha$ gets smaller, there is initially an increase in the $u_x$ component since pressure on the surface of the droplet is spread increasingly more onto a surface element that points in the x-direction. However, as $\alpha$ decreases more this surface element gets smaller too and eventually disappears completely as $\alpha \to 0$. Therefore, there is a competition between the decreasing area in this surface element and the increasing direction of the normal pointing more towards the x-direction. Hence, we find a maximum in the tilt angle sensitivity for small $\alpha \approx 0.3$. However, we note that the model is not applicable for $\alpha \ll 1$, since nonlinear plasma and fluid dynamics effects become increasingly more important when all laser energy is focused into a tight spot. In that case, the complete plasma and droplet fluid dynamics must be taken into account. In practice, for the $\mu$m-sized droplets considered here such tight focus cannot be reached and thus typically $\alpha \gg 0$.

\section{Discussion and Industrial application}
Careful control over the tilt angle sensitivity and target expansion is of crucial importance for the operating stability and CE of EUV light sources\cite{Benschop2008,Banine2011, Fomenkov2017, Kawasuji2017}. In the following, we apply our now validated model to predict sensitivities for several practical cases that are immediately relevant for current state-of-the-art industrial droplet-based EUV light sources. In the industrial context, tilt sensitivity is typically expressed as $\theta_{\text{tilt}}/ \Delta x$ (in degrees tilt / $\mu$m misalignment). Following Eq.  \eqref{eq:tiltangletheoreticalexpaned},  $\theta_{\text{tilt}}/ \Delta x = f(\alpha)/R_{\text{eff}}$.
 
 \begin{figure}[b!]
 \includegraphics[width=8.5cm]{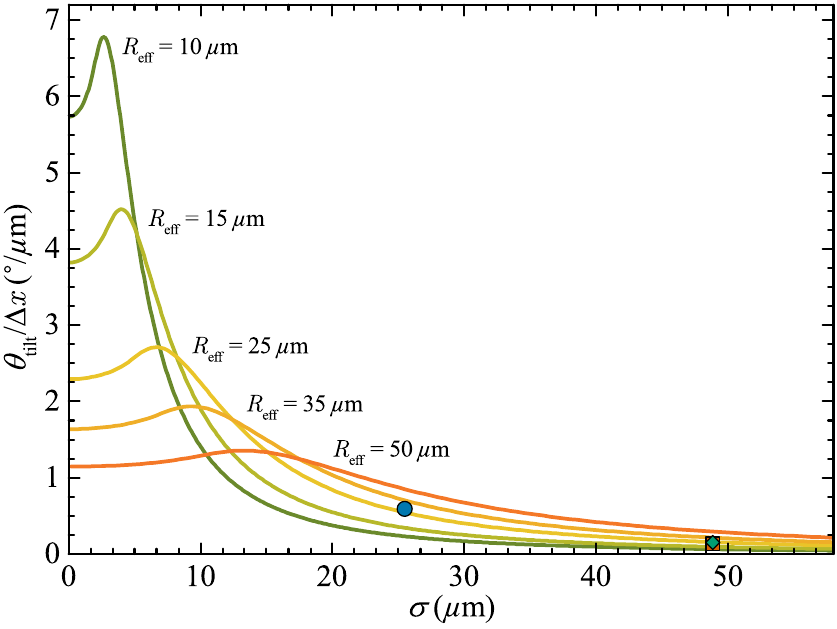}
\caption{\label{fig:tiltanglesensitivityexample} The calculated tilt angle sensitivity $f(\alpha)/R_\textrm{eff}$ expressed in units $^\circ / \mu$m as function of beam width $\sigma$ around zero misalignment ($\beta = 0$).  The curves are the analytically calculated tilt angle sensitivities from Eq.~\eqref{eq:tiltangletheoretical}, with $\delta_p = 0.6$ and for several relevant values for the effective droplet size $R_\textrm{eff}=[10,15,25,35,50]$\,$\mu$m. The red square, green diamond and blue circle represent the values corresponding to the experimental conditions (see Section~\ref{sec:RES}, color coding as in Fig.~\ref{fig:results}).}
\end{figure} 
 
In Fig.~\ref{fig:tiltanglesensitivityexample} we present our model predictions for the sensitivity $\theta_{\text{tilt}}/ \Delta x$ as function of beam width $\sigma$ around zero misalignment ($\Delta x = 0$) for several values of the effective droplet size $R_\textrm{eff} = [10,15,25,35,50]$ $\mu$m. We note that by plotting for several effective droplet sizes $R_{\textrm{eff}}$, we incorporate both the droplet size $R_0$ and the distance from the droplet surface to the plasma critical surface $d_{\text{crit}}$ for each case, since $R_\textrm{eff} = R_0 + d_\textrm{crit}$ (see previous section). In the figure we also show the experimental parameters studied in the previous sections, analogous to Fig.~\ref{fig:tiltanglesensitivity}.

Figure \ref{fig:tiltanglesensitivityexample} shows that increasing the laser spot size beyond $\sim$45\,$\mu$m does not significantly change the sensitivity for the given effective droplet sizes and is therefore not useful. However by increasing the laser spot size, the energy required to maintain a similar droplet expansion increases with $1/\sigma^2$ due to finite overlap between the droplet and the laser beam, as discussed in Ref.~[\onlinecite{Kurilovich2016}]. Therefore in practical applications, one should find the optimum conditions balancing between a maximum expansion (i.e. minimizing $\sigma$) and a minimal tilt sensitivity (i.e. maximizing $\sigma$).

Furthermore, from Fig.~\ref{fig:tiltanglesensitivityexample} we observe that the tilt angle sensitivity increases sharply with decreasing laser focus spot size, especially when $\sigma < R_\text{eff}$. Under such focusing conditions, a change in the effective size of the droplet $R_{\text{eff}}$ has a strong effect on the sensitivity. An interesting way to change the tilt sensitivity, apart from adapting the actual droplet radius $R_0$, is by changing the laser pulse energy or its wavelength. Shorter wavelengths or lower pulse energies result in a smaller $d_{\text{crit}}$ and hence result in a smaller $R_{\text{eff}}$ and vice versa. 

A particularly interesting industrial application of the model is found in the use of a nanosecond-long CO$_2$-laser prepulse, at 10.6-$\mu$m wavelength. According to radiation-hydrodynamics simulations\cite{Basko2015} of the interaction of such energetic laser pulses with tin droplets (at an absorbed intensity of $4$ x $10^9$ W/cm$^2$), the critical surface extends up to about 28\,$\mu$m from the droplet surface. In the particular case of $R_0=15\,\mu$m tin droplets impacted by a laser beam of $\sigma= 25\,\mu$m studied in Ref.~[\onlinecite{Basko2015}], we speculate that the effective system size in our model $R_\text{eff} \approx 15\,\mu\text{m} + 28\,\mu\text{m} \approx 43\,\mu\text{m}$. In addition, recent experimental work using CO$_2$-lasers impinging on planar solid tin targets showed that the exponent in Eq.\,(\ref{eq:criticalsurfacescaling}) is significantly larger, $\delta=0.96$, compared to the case of a Nd-YAG laser\cite{1612-202X-15-1-016003}. These results allows us to estimate the scaling of the propulsion velocity (and thus, radial expansion velocity\cite{Gelderblom2016}) with CO$_2$-laser intensity for the $\sigma= 25\,\mu$m case. Using these values for $R_\text{eff}$ and $\delta$ in Eq.\,(\ref{eq:tiltanglesensitivityappendix}), we get a sensitivity around zero misalignment of $\theta/ \Delta x \approx 0.98^\circ / \mu$m, which is about 85\% larger than the corresponding sensitivity for Nd:YAG $\theta/ \Delta x \approx 0.53^\circ / \mu$m. For the low-energy Nd:YAG cases studied in this work, we found that our experiments were well reproduced assuming $R_\text{eff} \approx R_0$, which would for the current example lead to a sensitivity of $0.36^\circ / \mu$m, which differs from the CO$_2$ case by a factor of 2.7. Of course, the extrapolation of our model to other laser wavelengths requires further experimental validation and is left for future work.

Certain industrial applications may require a finite tilt angle\cite{rafac2016}. Our model Eq.~\eqref{eq:tiltangledefinition} offers a direct way to predict what misalignment is required to obtain a certain amount of target tilt (see also Fig. \ref{fig:results}a). The slope of that curve around the required misalignment then gives the new tilt angle sensitivity, which can be calculated numerically by evaluating Eq.~\eqref{eq:tiltangletheoretical} around this new working point. We note that there are in-fact two planes in which we can induce a finite tilt angle, namely in the $x-z$ plane (as discussed in this work) but also in the $y-z$ plane. Both angles are dependent on misalignments in both $x$ and $y$ directions when the degeneracy is lifted by choosing a finite target tilt through a well-defined, intentional misalignment. Further study is required to see how this fact may be advantageously used to increase source operating stability by minimizing L2D sensitivity along the machine axis with the largest risk of misalignment.

\section{\label{sec:diss} Conclusions}
In tin-droplet-based LPP sources of EUV light, laser-to-droplet alignment plays an important role. A slightly misaligned prepulse laser beam can lead to a non-optimal target shape, which causes an inefficient coupling with the main laser pulse and lower conversion efficiency of drive laser light into EUV. Moreover, reflections of the main pulse laser light from the tilted surface may well be detrimental to laser stability. 

In this work we experimentally validated a simple, intuitive model describing the tilt angle sensitivity of a droplet impacted by a laser pulse with controlled misalignment. Our back-of-the-envelope model for the tilt angle was derived based solely on the direction of the center-of-mass velocity. From this model, we were able to obtain the local tilt angle sensitivity around zero misalignment. We experimentally verified the tilt angle and the tilt angle sensitivity by three different experimental groups with industrially relevant settings of an Nd:YAG laser operating at its fundamental wavelength. We observed an excellent agreement with the model over a broad range of laser pulse energies and two laser focus spot sizes. Further, we applied our validated model to predict sensitivities for several practical cases that are immediately relevant for current state-of-the-art industrial droplet-based EUV light sources.

Our model is a simple first-order approximation of the underlying plasma and fluid physics. Full three-dimensional simulations incorporating the complete plasma dynamics should be carried out to obtain the tilt angle as function of the full parameter space. Nonetheless, the current model already allows to physically understand the target tilt as a function of the key experimental control parameters.

\begin{acknowledgments}
We thank Alexander Klein for fruitful discussions. This work is part of an Industrial Partnership Programme of the Netherlands Organization for Scientific Research (NWO). This research programme is co-financed by ASML. Part of this work has been carried out at the Advanced Research Center for Nanolithography (ARCNL), a public-private partnership between the University of Amsterdam (UvA), the Vrije Universiteit Amsterdam (VU), NWO and ASML.
\end{acknowledgments}

\section*{Appendix}
We present the full solution of the tilt angle sensitivity $f(\alpha,\delta)$ around $\beta=0$ following (\ref{eq:tiltangletheoretical}) as used in Fig.~\ref{fig:tiltanglesensitivity},

\begin{widetext}
\begin{eqnarray}
f(\alpha,\delta)&=&2^{-\frac{\delta }{2}-3} e^{-\frac{1}{2} i \pi  \delta } \delta ^{\frac{1}{2}-\frac{\delta }{2}} \left(-\frac{\delta }{\alpha ^2}\right)^{\delta /2} \Gamma
   \left(\frac{\delta +1}{2}\right) \nonumber\\
   &\times&\frac{\left(-2 \delta ^{\frac{\delta +3}{2}} (\delta +3) e^{\frac{1}{2} \left(\frac{1}{\alpha ^2}+i \pi \right) \delta }-i 2^{\frac{\delta
   +3}{2}} \left(\alpha ^2 (\delta +1)+\delta \right) \alpha ^{\delta +1} \left((\delta +3) \Gamma \left(\frac{\delta +3}{2},-\frac{\delta }{2 \alpha ^2}\right)-2 \Gamma
   \left(\frac{\delta +5}{2}\right)\right)\right)}{\alpha ^4 \Gamma \left(\frac{\delta +5}{2}\right) \left(\delta  \Gamma \left(\frac{\delta }{2}\right)-2 \Gamma
   \left(\frac{\delta }{2}+1,-\frac{\delta }{2 \alpha ^2}\right)\right)}.
   \label{eq:tiltanglesensitivityappendix}
\end{eqnarray}
\end{widetext}

\bibliography{mybib}
\end{document}